\documentclass[aps,twocolumn,showpacs,floatfix]{revtex4}
\usepackage{amsmath}
\usepackage{amsfonts}
\usepackage{amssymb}
\usepackage{epsfig}
\usepackage{graphicx}

\begin{document}
\title{The Electromagnetically Induced Transparency in Mechanical Effects of Light}
\author{G. S. Agarwal and Sumei Huang}
\affiliation{Department of Physics, Oklahoma State University,
Stillwater, Oklahoma 74078, USA}

\date{\today}

\begin{abstract}
We consider the dynamical behavior of a nanomechanical mirror in a
high-quality cavity under the action of a coupling laser and a probe
laser. We demonstrate the existence of the analog of
electromagnetically induced transparency (EIT) in the output field at the
probe frequency. Our calculations show explicitly the origin of EIT-like dips as well as the characteristic changes in dispersion from
anomalous to normal in the range where EIT dips occur. Remarkably
the pump-probe response for the optomechanical system shares all the
features of the $\Lambda$ system as discovered by Harris and
collaborators.

\end{abstract}
\pacs{42.50.Gy,42.50.Wk} \maketitle

\renewcommand{\thesection}{\Roman{section}}
\setcounter{section}{0}

\renewcommand{\baselinestretch}{1}\small\normalsize
Since its original discovery in the
context of atomic vapors, electromagnetically induced transparency (EIT) 
\cite{Harris1,Field,Boller} has been at the center of many important
developments in optical physics \cite{Fleischhauer} and has led to
many different applications, most notably in the context of slow
light \cite{Hau,Harris,Kash} and the production of giant
nonlinear effects. EIT is helping the progress towards studying
nonlinear optics at the single-photon level. EIT has been reported
in many other systems \cite{Ham}. More recently, EIT has been
discovered in meta materials \cite{Zhang,Papasimakis,Tassin,Chiam}
where resonant structures can be fabricated to correspond to dark
and bright modes. Resonators provide certain advantages
\cite{Smith} because by design we can manipulate EIT to produce
desired transmission properties of a structure. It would thus be
especially interesting to study resonators coupled to other systems
such as cavity optomechanical systems. Such nanomechanical systems
have attracted considerable interest recently
\cite{Genes,Favero,Sumei,Meystre,Schliesser,Hartmann,Aspelmeyer,Kippenberg}.
In this letter, we demonstrate the possibility of EIT in the context
of cavity optomechanics.

Before discussing our model and results, we set the stage for EIT in
cavity optomechanics. As in typical EIT experiments
\cite{Harris1,Field,Boller,Fleischhauer}, for example, in the context of
atomic vapors, we need to examine the pump-probe response of a nanomechanical oscillator of frequency $\omega_{m}$ coupled to a high-quality cavity via radiation pressure effects
\cite{Braginsky1,Braginsky2} as schematically shown in Fig. 1.
Thus, the cavity oscillator of frequency $\omega_{0}$ and the nano-oscillator interact nonlinearly with each other. The system
is driven by a strong pump field of frequency $\omega_{c}$. This
is the coupling field. The probe field has frequency
$\omega_{p}$ and is much weaker than the pump field. The
mechanical oscillator's damping is much smaller than that of the cavity
oscillator. This is very important for considerations of EIT.
The decay rate of the mechanical oscillator plays the same role
as the decay rate of the ground-state coherence in EIT experiments.
The analog of the two-photon resonance condition where EIT occurs
would be $\omega_{c}+\omega_{m}=\omega_{p}$. We show how the
absorptive and dispersive responses of the probe change by the
coupling field and how EIT emerges. We present a clear physical
origin of EIT in such a system.

Let us denote the cavity annihilation (creation) operator by $c$
($c^{\dag}$) with the commutation relation $[c,c^{\dag}]=1$. The
momentum and position operators of the nanomechanical oscillator
with mass $m$ are represented by $p$ and $q$. We also introduce the
amplitudes of the pump field and the probe field inside the cavity
$\varepsilon_{c}=\sqrt{2\kappa \wp_{c}/(\hbar\omega_{c})}$ and
$\varepsilon_{p}=\sqrt{2\kappa \wp_{p}/(\hbar\omega_{p})}$, where
$\wp_{c}$ is the pump power, $\wp_{p}$ is the power of the probe
field, and $\kappa$ is the cavity decay rate. Note that
$\varepsilon_{c}$ and $\varepsilon_{p}$ have dimensions of
frequency. The optomechanical coupling between the cavity field and
the movable mirror can be described by the coupling constant
$\chi_{0}=\hbar\omega_{0}/L$, where $L$ is the cavity length. The
Hamiltonian describing the whole system reads
\begin{figure}[!h]
\begin{center}
\scalebox{0.75}{\includegraphics{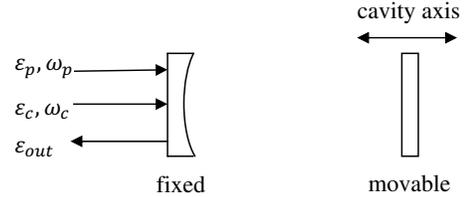}} \caption{\label{Fig1} Sketch
of the optomechanical system coupled to a high-quality cavity via
radiation pressure effects.}
\end{center}
\end{figure}
\begin{equation}\label{1}
\begin{array}{lcl}
\displaystyle H=\hbar\omega_{0}c^{\dag}c
+\left(\frac{p^2}{2m}+\frac{1}{2}m\omega_{m}^{2}q^{2}\right)+i\hbar\varepsilon_{c}(c^{\dag}e^{-i\omega_{c}t}-ce^{i\omega_{c}t})\vspace{0.1in}\\\hspace{0.3in}+i\hbar(c^{\dag}\varepsilon_{p}
e^{-i\omega_{p}t}-c\varepsilon_{p}^{*} e^{i\omega_{p}t})-\chi_{0}
c^{\dag}cq.
\end{array}
\end{equation}
This letter deals with the mean response of the system to the probe
field in the presence of the coupling field. Because we deal with the
mean response of the system we do not include quantum fluctuations.
This is similar to what has been done in the context of EIT work
where one uses atomic mean value equations and all quantum
fluctuations (due to either spontaneous emission or 
collisions) are ignored. Thus, we examine the mean value equations,
which can be obtained from the Hamiltonian and by addition of the
damping terms. We use the factorization assumption $\langle
Qc\rangle=\langle Q\rangle \langle c\rangle$ and also transform the
cavity field to a rotating frame at the frequency $\omega_{c}$, $
\langle c(t)\rangle=\langle\tilde{c}(t)\rangle e^{-i\omega_{c}t}$.
The mean value equations are then given by
\begin{equation}\label{2}
\begin{array}{lcl}
\displaystyle \langle\dot{q}\rangle=\frac{\langle p\rangle}{m},\vspace*{.1in}\\
\langle\dot{p}\rangle=-m\omega_{m}^2\langle q\rangle+\chi_{0} \langle \tilde{c}^{\dag}\rangle\langle\tilde{c}\rangle-\gamma_{m}\langle p\rangle,\vspace*{.1in}\\
\displaystyle\langle\dot{\tilde{c}}\rangle=-\left[\kappa+i\left(\omega_{0}-\omega_{c}-\frac{\chi_{0}}{\hbar}
\langle q\rangle\right)\right]\langle
\tilde{c}\rangle+\varepsilon_{c}+\varepsilon
_{p}e^{-i(\omega_{p}-\omega_{c})t}.
\end{array}
\end{equation}
The output field can be obtained by using the input-output relations
\cite{Walls}
\begin{equation}\label{3}
\begin{array}{lcl}
\displaystyle
\varepsilon_{out}(t)+\varepsilon_{p}e^{-i\omega_{p}t}+\varepsilon_{c}e^{-i\omega_{c}t}=2\kappa\langle
c\rangle.
\end{array}
\end{equation}
We first note that in the absence of the coupling field, the output
field is given by
\begin{equation}\label{4}
\begin{array}{lcl}
\displaystyle
\varepsilon_{out}(t)+\varepsilon_{p}e^{-i\omega_{p}t}=\varepsilon_{T}\varepsilon_{p}e^{-i\omega_{p}t}=\frac{2\kappa}{\kappa-i(\omega_{p}-\omega_{0})}\varepsilon_{p}e^{-i\omega_{p}t}.
\end{array}
\end{equation}
The quadratures of the field $\varepsilon_{T}$, defined by
$\varepsilon_{T}=\upsilon_{p}+i\tilde{\upsilon}_{p}$, show the
absorptive and dispersive behavior as a function of the detuning
parameter $(\omega_{p}-\omega_{0})$. The field quadratures, as is well
known, can be measured by homodyne techniques \cite{Walls}.

Next, we examine the effect of the coupling field. Equations
(\ref{2}) are nonlinear, and therefore the steady-state response contains many Fourier components. We solve in the limit of
arbitrary strength of the coupling field; however, we take the probe
field to be weak. We specifically are interested in the
response of the cavity optomechanical system to the probe in the
presence of the coupling field $\varepsilon_{c}$. Thus, we find the
component of the output field oscillating at the probe frequency
$\omega_{p}$. The result of such a calculation is that
$\varepsilon_{T}$ is now given by
\begin{equation}\label{5}
\begin{array}{lcl}
\displaystyle
\varepsilon_{T}=\frac{2\kappa}{d(\delta)}\{(\delta^2-\omega_{m}^2+i\gamma_{m}\delta)[\kappa-i(\Delta+\delta)]-2i\omega_{m}\beta\},
\end{array}
\end{equation}
where
 \begin{equation}\label{6}
\begin{array}{lcl}
\displaystyle
d(\delta)=(\delta^{2}-\omega_{m}^2+i\gamma_{m}\delta)[(\kappa-i\delta)^2+\Delta^2)]+4\Delta\omega_{m}\beta,\vspace{0.1in}\\
\displaystyle \delta=\omega_{p}-\omega_{c},\vspace{0.1in}\\
\displaystyle \Delta=\omega_{0}-\omega_{c}-\frac{2\beta\chi_{0}}{\omega_{m}},\vspace{0.1in}\\
\displaystyle \beta=\frac{\chi_{0}^{2}|\tilde{c}_{0}|^2}{2m\hbar\omega_{m}},\vspace{0.1in}\\
\displaystyle \tilde{c}_{0}=\frac{\varepsilon_{c}}{\kappa+i\Delta}.
\end{array}
\end{equation}
The coupling field has modified the output field at the probe
frequency. Note that $\varepsilon_{T}$ is nonperturbative in terms
of the strength of the coupling field $\omega_{c}$. We concentrate
on the output field. However, all the results for $\varepsilon_{T}$
also apply to the cavity field at $\omega_{p}$ as the two quantities
are proportional to each other.

In order to understand the coupling-field-induced modification of
the probe response $\varepsilon_{T}$, we make reasonable
approximations. We work in the sideband resolved limit
$\omega_{m}\gg\kappa$. This is the limit in which normal mode
splitting \cite{Marquardt,Kippenberg,Aspelmeyer} has been
discovered. Because it is known that the coupling between the nano-oscillator and the cavity is strongest whenever
$\delta=\pm\omega_{m}$ or
 $\delta=\pm\Delta$, the case $\Delta\sim\omega_{m}$ is considered here. After some
simplifications, we can write the output field in an instructive
form,
\begin{equation}\label{7}
\begin{array}{lcl}
\displaystyle
\varepsilon_{T}=\upsilon_{p}+i\tilde{\upsilon}_{p}=\frac{2\kappa}{\kappa-ix+\displaystyle\frac{\beta}{\frac{\gamma_{m}}{2}-ix}}=\frac{A_{+}}{x-x_{+}}+\frac{A_{-}}{x-x_{-}},
\end{array}
\end{equation}
where $x=\delta-\omega_{m}$, which is the detuning from the line
center. Further, it is seen that the denominator has two roots,
which are
\begin{equation}\label{8}
\begin{array}{lcl}
\displaystyle
x_{\pm}=\frac{-i(\kappa+\frac{\gamma_{m}}{2})\pm\sqrt{-(\kappa-\frac{\gamma_{m}}{2})^2+4\beta}}{2},
\end{array}
\end{equation}
whose nature depends on the power of the coupling laser. For
coupling powers less than the critical power
\begin{equation}\label{9}
\begin{array}{lcl}
\displaystyle
\tilde{\wp}_{c}=\frac{\hbar\omega_{c}|\tilde{c}_{0}|^2(\kappa^2+\omega_{m}^2)(\kappa-\frac{\gamma_{m}}{2})^2}{8\kappa\beta},
\end{array}
\end{equation}
the two roots are purely imaginary. For $\wp_{c}>\tilde{\wp}_{c}$,
the roots are complex conjugates of each other. The region
$\wp_{c}>\tilde{\wp}_{c}$ corresponds to the region where normal-mode splitting \cite{Marquardt,Kippenberg,Aspelmeyer} occurs and has
been studied recently using a very different technique. In the
context of optical physics, this is the region where Autler-Townes
splitting \cite{Autler} occurs, although sometimes the distinction
between different kinds of splittings is marred. However, for EIT, it
is important to have $\gamma_{m}\ll\kappa$.

\begin{figure}[!h]
\begin{center}
\scalebox{0.8}{\includegraphics{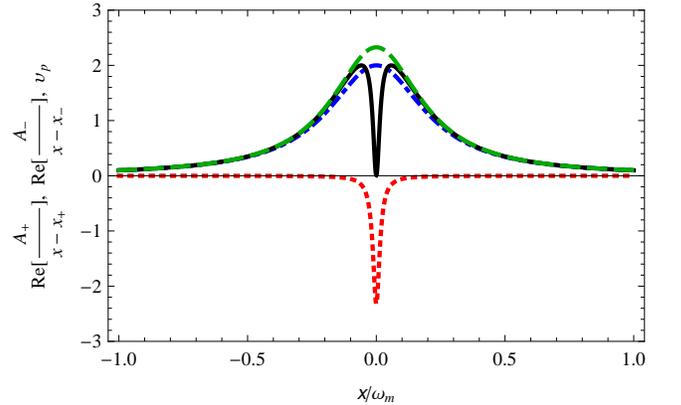}} \caption{\label{Fig2}(Color
online) Quadrature of the output field $\upsilon_{p}$ (solid
black curve) and the different contributions to it: the real parts
of $\frac{A_{+}}{x-x_{+}}$ (dotted red curve) and
$\frac{A_{-}}{x-x_{-}}$ (dashed green curve) as a function of the
normalized frequency $x/\omega_{m}$ for input coupling laser power
$\wp_{c}=1$ mW. The dot-dashed blue curve is $\upsilon_{p}$ in the
absence of the coupling laser.}
\end{center}
\end{figure}
\begin{figure}[!h]
\begin{center}
\scalebox{0.8}{\includegraphics{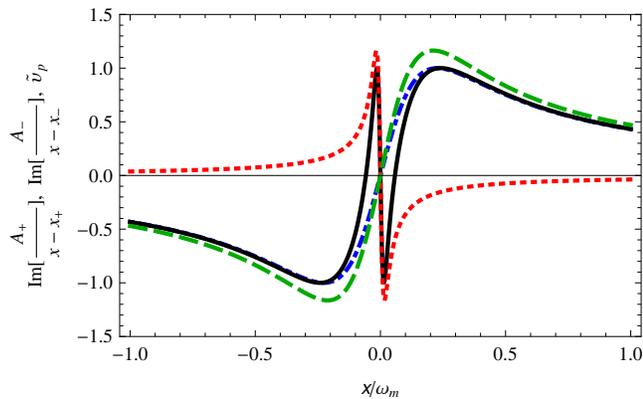}} \caption{\label{Fig3}(Color
online) Quadrature of the output field $\tilde{\upsilon}_{p}$
(solid black curve) and the different contributions to it: the
imaginary parts of $\frac{A_{+}}{x-x_{+}}$ (dotted red curve) and
$\frac{A_{-}}{x-x_{-}}$ (dashed green curve) as a function of the
normalized frequency $x/\omega_{m}$ for input coupling laser power
$\wp_{c}=1$ mW. The dot-dashed blue curve is $\tilde{\upsilon}_{p}$
in the absence of the coupling laser.}
\end{center}
\end{figure}

In order to bring out prominently features like EIT
\cite{Harris1,Field,Boller}, we specifically examine the case when
the coupling power is less than the critical power. Note that
$x_{+}\rightarrow-i\frac{\gamma_{m}}{2}$, $x_{-}\rightarrow-i\kappa$
as $\beta\rightarrow0$. Thus, the quadratures of the output field
have two distinct contributions in the limit of low values of the
coupling laser strength. One contribution is extremely narrow as
$\gamma_{m}\ll\kappa$. This characteristic property leads to the EIT
dip. For numerical work, we use parameters from a recent experiment
on the observation of the normal-mode splitting \cite{Aspelmeyer}:
the wavelength of the laser $\lambda=2\pi c/\omega_c=1064$ nm,
$L=25$ mm, $m=145$ ng, $\kappa=2\pi\times215$ kHz,
$\omega_m=2\pi\times947$ kHz, $\gamma_m=2\pi\times141$ Hz, the
mechanical quality factor $Q=\omega_{m}/\gamma_{m}=6700$. We calculate the critical power $\tilde{\wp}_{c}$ to be 3.8 mW.
In Figs. \ref{2} and \ref{3}, we show each contribution in Eq.
(\ref{7}) separately and also the total contribution. We observe
that the narrow contribution is inverted relative to the broad
contribution, and this leads to the typical EIT-like line shape for
the quadrature $\upsilon_{p}$ of the output field. The value at the
dip is not exactly zero as $\gamma_{m}\neq0$, though the value is
very small as $\gamma_{m}\ll\kappa$. This is similar to what one has
in the context of EIT in atomic systems where a strict zero is
obtained if the ground-state atomic coherence has an infinite lifetime. In the absence of the coupling field, the narrow feature
disappears (blue curve in the Fig. \ref{2}). The narrow feature's
width has a contribution which depends on the coupling laser power.
In leading order, the width is $\frac{\gamma_{m}}{2}+\frac{\beta}{\kappa}$. For the plot of the Fig.
\ref{3}, the power-dependent contribution to the width in
dimensionless units is $\beta/\kappa^2\sim0.065$. The quadrature
$\tilde{\upsilon}_{p}$ exhibits dispersive behavior, and the coupling
field changes the nature of dispersion from anomalous to normal in
the region where quantum interferences are prominent. This behavior
of dispersion is similar to the one found by Harris and
collaborators in predictions of slow light \cite{Hau,Harris,Kash} in
atomic systems.
\begin{figure}[!h]
\begin{center}
\scalebox{0.8}{\includegraphics{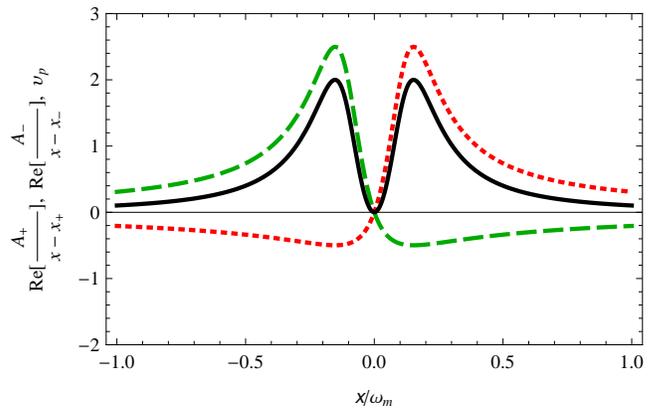}} \caption{\label{Fig4}(Color
online) Same as in Fig. 2 except the input coupling laser power
$\wp_{c}=6.9$ mW and $\wp_{c}=0$ case is not shown.}
\end{center}
\end{figure}
\begin{figure}[!h]
\begin{center}
\scalebox{0.8}{\includegraphics{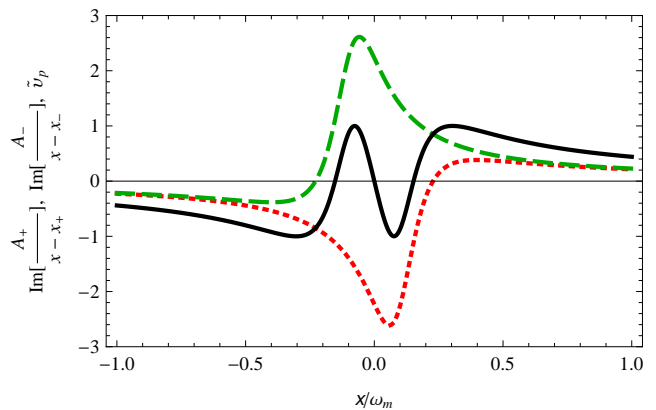}} \caption{\label{Fig5}(Color
online) Same as in Fig. 3 except the input coupling laser power
$\wp_{c}=6.9$ mW and $\wp_{c}=0$ case is not shown.}
\end{center}
\end{figure}

We next present the nature of interferences in the region when
$\wp_{c}>\tilde{\wp}_{c}$ in Figs. \ref{4} and \ref{5}. A
typical behavior is shown in Fig. \ref{4} which clearly shows
how the interference of the two contributions in Eq. (\ref{7}) leads
to the formation of the dip. The two contributions in Eq. (\ref{7})
lead to asymmetric profiles. In the region of EIT, the tails from
these contributions interfere. Unlike the case given by Fig.
\ref{2}, the two contributions have identical line widths. From
Fig. \ref{5}, we also see how the dispersive behavior is changed
by the coupling field from anomalous to normal in the region where
quantum interferences are dominated. The inverted nature of the
contribution $A_{+}$ should be noted, and it is this which changes
the nature of dispersion.

We now explain the origin of the structure (7) for the
probe response. Let us re-examine the Hamiltonian (1). Note that we
drive the cavity with arbitrary pump field $\varepsilon_{p}$. This
effectively prepares the cavity in a coherent state with a value
$\tilde{c}_{0}$ if all the other interactions were zero. The
trilinear interaction due to radiation pressure $\chi_{0}c^{\dag}cq$
can now be written as
$\chi_{0}q|\tilde{c}_{0}|^2+\chi_{0}q(\tilde{c}_{0}^{*}\delta
c+\tilde{c}_{0}\delta c^{\dag})+$ higher order terms if we write the
cavity operator $c$ as $\tilde{c}_{0}+\delta c$. The pump thus has
resulted in a bilinear interaction between the cavity oscillator and
the mirror oscillator. The cavity oscillator is driven by the probe
field, whereas the matter oscillator has no external drive. The
cavity oscillator is damped at the rate $\kappa$, whereas the mirror
is damped at the rate $\gamma_{m}\ll\kappa$. This situation
typically results \cite{Zhang, Papasimakis, Tassin, Chiam, Smith} in
line shapes such as (7).

In conclusion, we have shown how an exact analog of EIT can occur in cavity
optomechanics when such a system is driven by a weak probe in the presence of a strong coupling field. We find that the response
function for the cavity field at the probe frequency as well as the
output field has exactly the same features as the response of a
$\Lambda$ system provided the damping of the nanomechanical mirror is
much smaller than the dissipation in the cavity. We further
highlighted the interference effects in  two distinct regions of the
coupling power.

We gratefully acknowledge support from NSF Grant Phys. 0653494.

\end{document}